\documentclass[a4paper,10pt]{article}

\usepackage{amsfonts,amsthm,amssymb}
\usepackage{amsfonts}
\usepackage{amsmath}
\usepackage{amssymb}
\usepackage{graphicx}%\usepackage{epsfig}
\usepackage{dcolumn}
\usepackage{graphicx}
\usepackage[colorlinks,urlcolor=blue,citecolor=blue]{hyperref}
\begin{document}

\title{Spectrally isomorphic Dirac systems: graphene in electromagnetic field}
\author{{V\'i{}t Jakubsk\'y}\\
{\small \textit{
Department of Theoretical Physics,
Nuclear Physics Institute,
25068 \v Re\v z, Czech Republic }}\\
\sl{\small{E-mail:jakub@ujf.cas.cz
%\texttt{at} 
} }}
\date{}
\maketitle	
\begin{abstract}
We construct the new one-dimensional Dirac Hamiltonians that are spectrally isomorphic (not isospectral) with the known exactly solvable models. Explicit formulas for their spectra and eigenstates are provided. The operators are utilized for description of Dirac fermions in graphene in presence of an inhomogeneous electromagnetic field. We discuss explicit, physically relevant, examples of spectrally isomorphic systems with both non-periodic and periodic electromagnetic barriers. In the latter case, spectrally isomorphic two- and three-gap systems associated with the Ablowitz-Kaup-Newell-Segur hierarchy are considered.
\end{abstract}

\section{Introduction}

The effective one- or two-dimensional Dirac equation describes surprising variety of physical systems. It appears in effective models of spontaneous symmetry breaking in the quantum field theory \cite{NJL},  \cite{152}, \cite{519},  in the Andreev approximation of the Bogoliubov-de Gennes equations of superconductivity \cite{520}, or in the effective description of the low-energy charge carriers in graphene and related carbon nanostructures \cite{345}. In all these scenarios, the explicit solution of dynamical equations is desirable. However, it cannot be found rather frequently due to analytically intractable form of interaction. Here, the importance of exactly solvable models comes out. Despite their family is rather limited, they can be adjusted to the required physical situation by variety of methods.

Unitary transformations in quantum mechanics are convenient tool for making calculations more feasible, however, they are usually not considered to have much physical importance as the unitarily equivalent quantum systems provide the same physical predictions. Nonetheless, the unitary transformation can relate two systems whose physical interpretation is substantially different from each other. As an example, let us mention the free massless Dirac particle on the real line whose Hamiltonian reads $H=-i\sigma_1\partial_x$. The unitary transformation $U=e^{i\sigma_1 V_0(x)}$ gives rise to the operator
\begin{equation}
 \tilde{H}=UHU^{-1}=-i\sigma_1\partial_x+V_0(x)
\end{equation}
that can be interpreted as the Hamiltonian of Dirac fermion in carbon nanotube in presence of a scalar potential $V_0(x)$. The relation between $H$ and $\tilde{H}$ has a nontrivial physical impact here; it explains the absence of backscattering of Dirac quasiparticles on impurities in carbon nanotubes \cite{84}. 

Two quantum systems related by unitary transformation have the same energy spectrum; they are \textit{isospectral}. All the relevant properties of the transformed system can be effortlessly  obtained from the initial model. Similar advantage is offered by the supersymmetric quantum mechanics,  where the unitary equivalence is generalized by the intertwining relation $qH=\tilde{H}q$ and its conjugate. Instead of unitarity, the intertwining operator $q$, known also as Darboux transformation, satisfies $q^{\dagger}q=P(H)$ where $P$ is a (polynomial) function of the initial Hamiltonian \cite{529}. This transformation was discussed in \cite{474} for the one-dimensional Dirac operators where explicit formulas for $\tilde{H}$ satisfying the intertwining relations for a fixed $H$ and $q$ were provided. The scalar potential is invariant with respect to the Darboux transformation $q$, i.e. it has the same form in both $H$ and $\tilde{H}$. The spectra of $\tilde{H}$ and $H$ are identical up to a finite number of energy 
levels; the systems are \textit{almost isospectral}.  The  Dirac Hamiltonians $H$ and $\tilde{H}$ intertwined by $q$ describe different interactions in general. For instance, $\tilde{H}$ can represent the Hamiltonian of radially twisted carbon 
nanotubes whereas $H$ is associated with a twist-free nanotubes, see \cite{85}.

We aim to construct the new, solvable, one-dimensional Dirac Hamiltonian that would describe Dirac fermions in graphene in presence of an inhomogeneous electromagnetic field. The procedure should allow us to modify the scalar potential and to introduce or maintain a constant mass term. The initial and the transformed Hamiltonians do not need to be isospectral. We are rather interested in the systems where the isospectrality of $H$ and $\tilde{H}$ ($E\in\sigma(H)\Leftrightarrow E\in\sigma(\tilde{H})$) is generalized by the \textit{spectral isomorphism}, defined in terms of a real function $f$,
\begin{equation}\label{isomorph}
 E\in\sigma(H)\Leftrightarrow f(E)\in\sigma(\tilde{H}).
\end{equation}
The construction will be divided into two independent steps. The first step comprises of the special unitary transformation whereas  the second one allows us to modify the spectrum of the operator globally, introducing spectrally isomorphic systems in a simple way.

Dirac fermion in graphene is described by the Dirac equation \cite{345}
\begin{equation}\label{h}
\left(\sum_{a=1,2}\sigma_a\left[v_F(-i\hbar \partial_a+e A^{mg}_a)+A^{d}_a\right]+m \sigma_3+V \right)\Psi=E\Psi, 
\end{equation}
where $\mathbf{A}^{mg}$ is the vector potential corresponding to the external magnetic field whereas $\mathbf{A}^d$ corresponds to the pseudo-magnetic field induced by mechanical deformation \cite{97}. The pseudo-spin distinguishes two triangular sublattices of the hexagonal lattice. The mass term $m$ makes the two sublattices inequivalent; it is absent in the suspended graphene but can appear when the graphene sheet is posed on a substrate, e.g. on the crystal of hexagonal boron-nitride \cite{10}.  

We will focus on the settings with translational invariance in one direction. We suppose that the pseudo-magnetic and magnetic field have nonvanishing $z$-component only, 
$ B_z=B_z\left(x_1\right)=\partial_{1}A_2,$ $A_2=A_2(x_1)$, $A_1=0$, $m=const.\geq0$, $V=V(x_1)$.
For convenience, we will work with the rescaled coordinates
$ x_1=\lambda x$, $x_2=\lambda y$,
where $x$ and $y$ are dimensionless quantities and $\lambda$ has dimension of length (it can be identified with the magnetic length $l_B=\sqrt{\frac{\hbar}{e B_0}}$ for instance).
After the substitution and separation of variables $\psi(x,y)=e^{i k_y y}\psi(x)$, the equation (\ref{h}) can be written as
\begin{equation}\label{tildeHH}
 \tilde{H}\tilde{\psi}=\left(-i\sigma_1\partial_x+\tilde{V}_2\sigma_2+\tilde{m}\sigma_3+\tilde{V}_0\right)\psi(x)=\tilde{E}\psi(x),
\end{equation}
where
$\tilde{V}_0=\frac{\lambda}{\hbar v_F}V_0(\lambda x),$ $\tilde{E}=\frac{\lambda E}{v_F \hbar}$, $
\tilde{V}_2=\lambda \left(\frac{e}{\hbar}A_2^{mg}(\lambda x)+\frac{1}{v_F\hbar}A_2^{d}(\lambda x)\right)+k_y$, and $\tilde{m}=\frac{\lambda}{v_F\hbar}m$  
are dimensionless quantities. We used  suggestively the tilded notation in (\ref{tildeHH}) as we shall construct solvable Hamiltonians of this type.

\section{The special unitary transformation}
Let us take a generic one-dimensional Dirac operator $H$, 
\begin{equation}
 H=-i\sigma_1\partial_x+V_2(x)\sigma_2+V_3(x)\sigma_3+V_0(x),
\end{equation}
with real, sufficiently smooth, functions $V_0(x)$, $V_2(x)$ and $V_3(x)$, where $V_2(x)$ and $V_3(x)$ do not vanish simultaneously. We are looking for a ``gauge'' transformation $G$ that would change $H$ to $\tilde{H}$ of (\ref{tildeHH}),
\begin{equation}\label{tildeH}
 \tilde{H}=GHG^{-1}=-i\sigma_1\partial_x+\tilde{V}_2\sigma_2+\tilde{m}\sigma_3+\tilde{V}_0.
\end{equation}
We take $G$ as an element of the $su(2)$ group,
\begin{equation}\label{generalG}
 G=g_0+i\sum_{j=1}^3 g_j\sigma_j,\quad g_0^2+\sum_{j=1}^3g_j^2=1,
\end{equation}
where $g_a$, $a=0,1,2,3$ are real and continuous functions.
Transformation of the kinetic term of $H$ results in the following expression
\begin{equation}
 i G\sigma_1\partial_xG^{\dagger}=i\left[(g_0^2+g_1^2-g_2^2-g_3^2)\sigma_1+2(g_1g_2-g_0g_3)\sigma_2+2(g_1g_3+g_0g_2)\sigma_3\right]\partial_{x}+\dots.,
\end{equation}
where the dots represent the terms without derivative. We require the coefficient at the derivative to be constant\footnote{Position dependent coefficient at the derivative corresponds to the position dependent Fermi velocity. These systems were considered in the literature, see e.g. \cite{250}.   
}. It can be satisfied by the following simple ansatz for $G$,
\begin{equation}\label{G}
 G(x)=g_0(x)+i\sigma_1 g_1(x),\quad  g_0^2+g_1^2=1.
\end{equation}
Substituting this expression into 
(\ref{tildeH}), we get
\begin{eqnarray}\label{potentials}
\tilde{V}_0&=&V_0+g_1g'_0-g_0g'_1,\\
\tilde{V}_2&=&(g_0^2-g_1^2)V_2+2g_0g_1V_3,\\
\tilde{m}&=&-2g_0g_1V_2+(g_0^2-g_1^2)V_3.
\end{eqnarray}
The mass term $\tilde{m}$ is required to be a non-negative constant. It can be achieved by $G\equiv G^{\pm}=g^{\pm}_0+i\sigma_1g^{\pm}_1$ where the functions $g_0^{\pm}$ and $g_1^{\pm}$ are fixed as  
\begin{equation}\label{gauge}
 g_0^{\pm}=\sqrt{\frac{1}{2}+\frac{\tilde{m} V_3\mp V_2\sqrt{V_3^2+V_2^2-\tilde{m}^2}}{2(V_2^2+V_3^2)}},\quad g_1^{\pm}=-g_0^{\pm}\frac{V_2\pm\sqrt{V_3^2+V_2^2-\tilde{m}^2}}{\tilde{m}+V_3}.
\end{equation}

\begin{figure}
\centering
\begin{tabular}{cc}
\includegraphics[scale=.6]{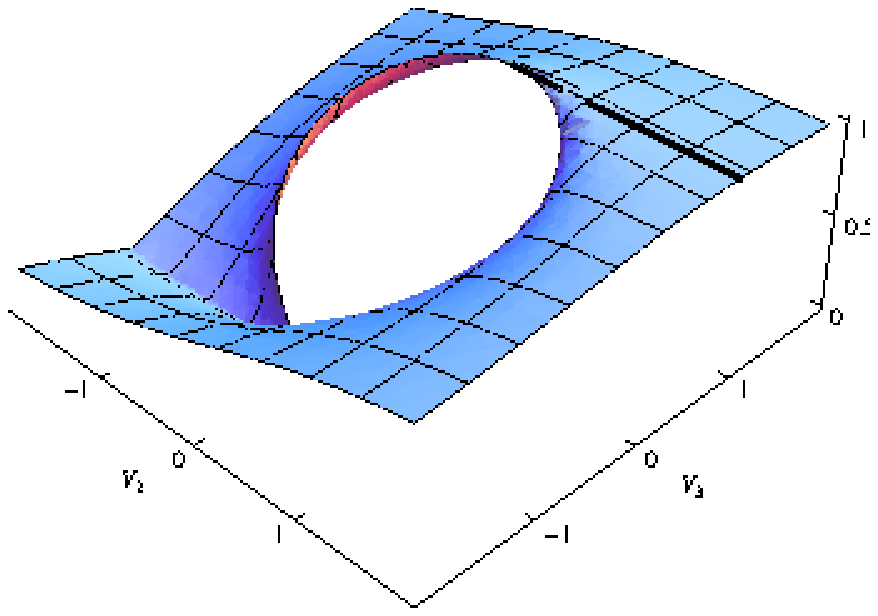}&
\includegraphics[scale=.6]{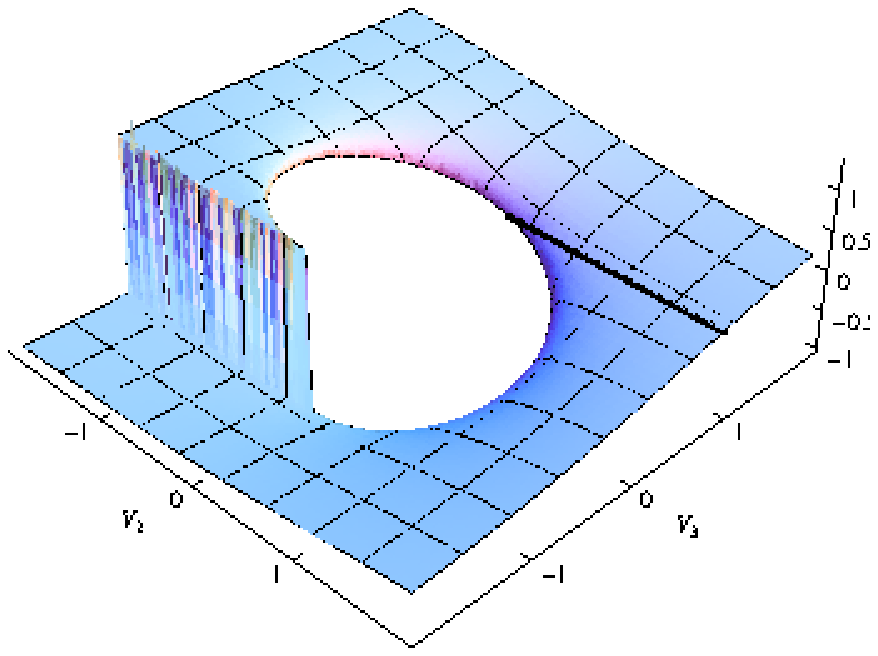}\\
\includegraphics[scale=.5]{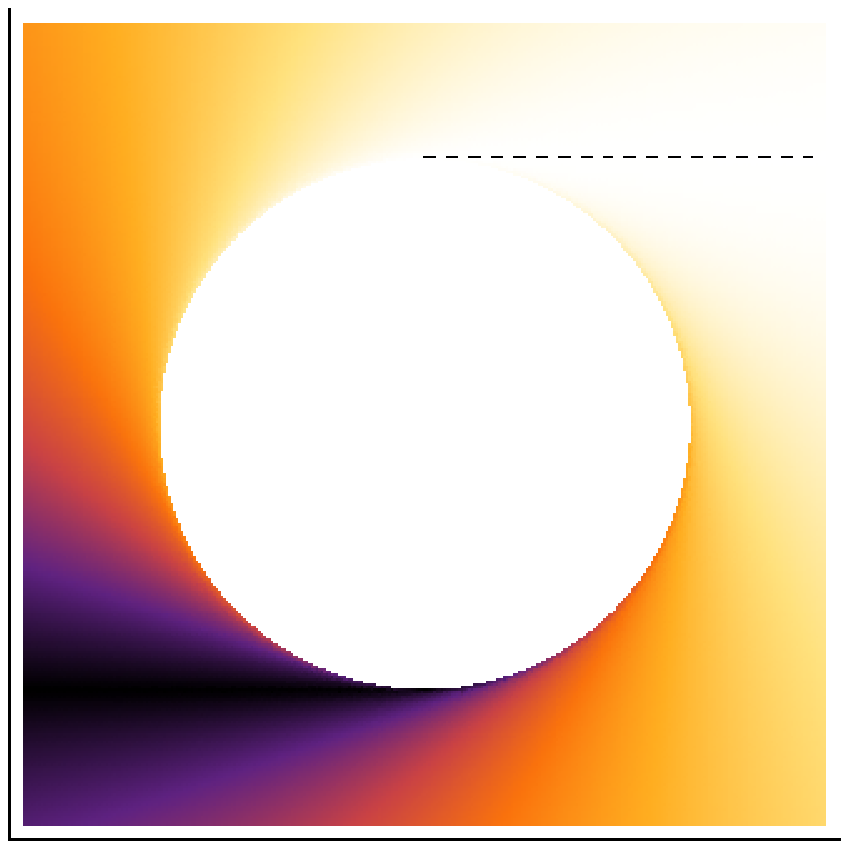}&
\includegraphics[scale=.5]{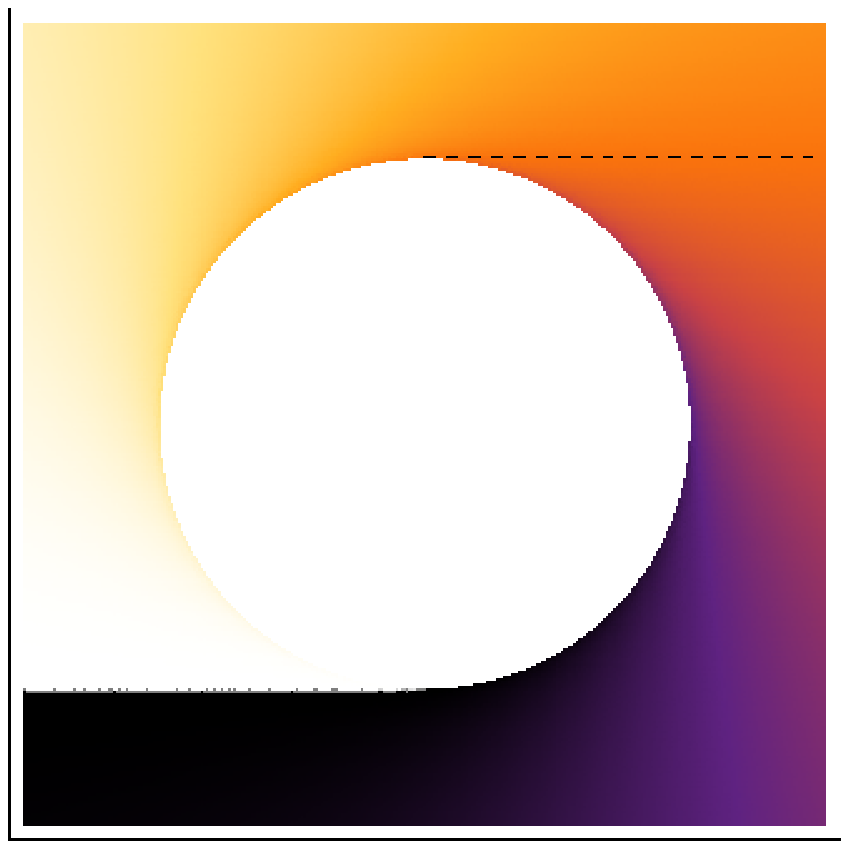}
\end{tabular}\caption{The graphs of $g_0^-$ (left) and $g_1^-$ (right). It is worth noticing that for $V_3=\tilde{m}$, there holds $g_0^{-}=1$ and $g_1^-=0$ for $V_2>0$. Hence, $G^-$ reduces to identity operator. $g_0^-=1$ and $g_1^-=0$ are represented by the black line in each figure. The density plots illustrate the circular domain of radius $\tilde{m}$ where $g_0^-$ and $g_1^-$ are imaginary. }\label{fig01}
\end{figure}

The functions (\ref{gauge}) have to be real. One can check\footnote{It is convenient to use $V_2=R\sin\phi$, $V_3=R\cos\phi$, $R\geq \tilde{m}$. Then the argument of the square root in $g_0^+$ is real if and only if $(R+\tilde{m}\cos\phi)\geq\sqrt{R^2-\tilde{m}^2}\sin\phi$. The left-hand side is always positive. Whenever $\sin\phi\leq 0$, the right-hand side is negative and the inequality is satisfied. For $\sin\phi>0$, the inequality is satisfied again as it reduces to $(R\cos\phi+\tilde{m})^2\geq0$. The same conclusion can be obtained for $g_0^-$.  } that this requirement is satisfied whenever $V_2^2+V_3^2>\tilde{m}^2$.
Besides, the gauge transformation $G$ has to be continuous. Otherwise, it would introduce discontinuities into the wave functions. As $V_2$ and $V_3$ are continuous, we can analyze the continuity of $g_0^{\pm}$ and $g_1^{\pm}$ in the $(V_2,V_3)$-plane, taking $V_2$ and $V_3$ as two independent variables. The function $g_1^{+}$ has discontinuity at  $V_3=-\tilde{m}$ for $V_2>0$ whereas $g_1^-$ has discontinuity at $V_3=-\tilde{m}$ for $V_2<0$, see Fig.\ref{fig01} for illustration. Hence, the continuous gauge transformation $G$ can be constructed as long as the values of $V_2$ and $V_3$ of the seed Hamiltonian avoid one of the indicated discontinuities at least.  Substituting (\ref{gauge}) into (\ref{potentials}), we get
\begin{eqnarray}\label{tildeV0}
 \tilde{V}_0^{\pm}&=&V_0+\frac{V_3V_2'-V_2V_3'}{2(V_2^2+V_3^2)}\pm\tilde{m}\frac{\left(\log (V_2^2+V_3^2)\right)'}{4\sqrt{V_2^2+V_3^2-m^2}},\\
 \tilde{V}_2^{\pm}&=&\mp\sqrt{-\tilde{m}^2+V_2^2+V_3^2},\label{tildeV2}\\
 \tilde{V}^{\pm}_3&=&\tilde{m}.
\end{eqnarray}

\section{Spectral isomorphism of massless and massive systems}
Let us discuss here a specific class of quantum systems whose Hamiltonian is obtained by addition of a mass term into a massless Dirac operator $H_0$ with a solvable vector potential, 
\begin{equation}
 H_{M}=H_0+M\sigma_3,\quad M>0,
\end{equation}
where 
\begin{equation}
H_0=-i\sigma_1\partial_x+V_2\sigma_2.
\end{equation}
We take for granted that the solutions of 
 $H_0\Psi^0_E=E\Psi^0_E$
are known. For $E\neq0$, the components of the spinor $\Psi^0_E=(\psi^0_E,\xi^0_E)^T$ satisfy the following equations:
\begin{equation}\label{psi0E}
 -{\psi^0_E}''+\left(V_2^2+V_2'-E^2\right)\psi^0_E=0,
\quad\xi^0_E=\frac{-i\partial_x+iV_2}{E}\psi^0_E. 
\end{equation}
The the zero modes of $H_0$ can be found easily,
\begin{equation}\label{zeromodes}
 \Psi_{+}^0=\left(e^{\int_{0}^x V_2(s)ds},0\right),\quad \Psi_{-}^0=\left(0,e^{-\int_{0}^x V_2(s)ds}\right), \quad H_0\Psi^{0}_{\pm}=0.
\end{equation}
Only one of them can be square integrable. Hence, if the normalizable zero-mode exists, it is non-degenerate. 

The mass term added to $H_0$ does not violate solvability. The stationary equation
\begin{equation}\label{stacM}
 (H_{0}+M\sigma_3)\Psi_{\epsilon}=\epsilon\,\Psi_{\epsilon} 
\end{equation}
is equivalent to the following system of differential equations,
\begin{equation}\label{s2}
 -\psi_{\epsilon}''+\left(V_2^2+V_2'-{\epsilon}^2+M^2\right)\psi_{\epsilon}=0,
\quad\xi_{\epsilon}=\frac{-\partial_x+V_2}{{\epsilon}+M}\psi_{\epsilon},\quad \Psi_{\epsilon}=(\psi_{\epsilon},\xi_{\epsilon})^T.
\end{equation}
The first one is identical to the equation for $\psi_E^0$ in (\ref{psi0E}) as long as we substitute $\epsilon^2=E^2+M^2$. Hence, $\psi_{\epsilon}=\psi^0_{E}$. Comparison of the second equation in (\ref{s2}) with the second one in (\ref{psi0E}) reveals that $\xi_{\epsilon}$ is proportional to $\xi^0_{E}$ up to a multiplicative constant. Hence, the relative coefficient between the spin-up and spin-down components of $\Psi_{\epsilon}$ gets altered. We get the following formula for the (non-normalized) eigenstates
\begin{equation}\label{psiM}
 \Psi_{\pm|\epsilon|}=\left(\begin{array}{cc}1&0\\0&\frac{|E|}{\sqrt{E^2+M^2}\pm M}\end{array}\right)\Psi_{\pm|E|}^0,\quad |\epsilon|=\sqrt{E^2+M^2},
\end{equation}
\begin{equation}
 H_M\Psi_{\pm|\epsilon|}=\pm|\epsilon|\Psi_{\pm|\epsilon|},\quad H_0\Psi_{\pm|E|}^0=\pm|E|\Psi_{\pm|E|}^0.
\end{equation}
The formula (\ref{psiM}) fails to provide solutions for $|\epsilon|=M$. We can find them by direct solution of $H_M\Psi_{\pm M}=\pm M\Psi_{\pm M}$.  
The eigenfunctions are
\begin{equation}\label{psipM}
 \Psi_M=\left(\begin{array}{l}
               2b Mi \, e^{\int^x_0V_2}\int_{0}^xe^{-2\int^s_0V_2}+a\, e^{\int^x_0V_2}\\
               b\, e^{-\int^x_0V_2}
              \end{array}\right)
\end{equation}
and
\begin{equation}\label{psimM}
 \Psi_{-M}=\left(\begin{array}{l}   
	     b\, e^{\int^x_0V_2}\\
              -2bMi\, e^{-\int^x_0V_2}\int_{0}^xe^{2\int^s_0V_2}+a  e^{-\int^x_0V_2}
                 \end{array}\right),
\end{equation}
where $a$ and $b$ are constants. It is worth emphasizing that for $a=1$ and $b=0$, the functions coincide with (\ref{zeromodes}), 
\begin{equation}\label{M}
 \Psi_{\pm M}|_{a=1,b=0}=\Psi^0_{\pm}.
\end{equation}
It suggests that whenever $H_0$ has a normalizable zero-mode $\Psi^0_+$ (or $\Psi_{-}^0$), then $H_M$ has a normalizable bound state $\Psi_M$  (or $\Psi_{-M}$) with energy $E=M$ (or $E=-M$). When $H_0$ has doubly degenerate zero mode (e.g. in the case of periodic systems), then both $\Psi_{\pm M}$ represent physical states of $H_M$ and $E=\pm M$ belong to the spectrum of $H_M$. 

We suppose that when $\Psi_E^0$ belongs to the domain of $H_0$, then $\Psi_{\epsilon}$ it is in the domain of $H_M$ as well. This is satisfied provided $H_M$ inherits the boundary conditions for $H_0$ that do not fix relative coefficient between the spin-up and spin-down components of the wave functions. When this is the case, there also holds the inverse implication: when $\Psi_{\epsilon}$ is eigenstate of $H_M$, then $\Psi_E$ is an eigenstate of $H_0$.
Then the Hamiltonians $H_0$ and $H_M$ are spectrally isomorphic. The isomorphism between the energies of the two operators reads 
\begin{equation}\label{Ee}
 f:\pm|E|\rightarrow \left\{\begin{array}{cc}\pm\sqrt{E^2+M^2}\quad\mbox{for}\quad E\neq0,\\ M\ \mbox{or}\ -M\quad\mbox{for}\quad E=0.\end{array}\right.
\end{equation}
The formula (\ref{Ee}) tells us that the difference between the energy levels of $H_0$ and $H_M$ goes to zero in the high-energy limit.

$H_M$ can be utilized as a suitable operator for application of the gauge transformation (\ref{tildeH}). We identify $H\equiv H_M$. As the mass term  corresponding to $V_3=M$ is constant, it is sufficient to require $|\tilde{m}|< M$ in order to avoid discontinuities at $g_1^{\pm}$ and keep the functions $g_0^{\pm}$ and $g_1^{\pm}$ real (see discussion below (\ref{gauge})).  We get the Hamiltonian $\tilde{H}$
\begin{equation}
 \tilde{H}=-i\sigma_1\partial_x+\tilde{V}_2\sigma_2+\tilde{m}\sigma_3+\tilde{V}_0
\end{equation}
with the following potentials
\begin{eqnarray}
 \tilde{V}_0^{\pm}=\frac{V_2'\left(M\pm\frac{\tilde{m}V_2}{\sqrt{-\tilde{m}^2+V_2^2+M^2}}\right)}{2(M^2+V_2^2)},\quad \tilde{V}_2^{\pm}=\mp \sqrt{-\tilde{m}^2+M^2+V_2^2},\quad 
\end{eqnarray}
where the superscript tells which one of $G^{\pm}=g_0^{\pm}+i\sigma_1g_1^{\pm}$ was used to transform the Hamiltonian.

\section{Examples}

In this section, we present several examples of spectrally isomorphic, exactly solvable models. We focus on both non-periodic and periodic quantum settings and present their spectrally isomorphic partners. In all cases, we discuss relevance of the models for description of Dirac fermions in graphene in presence of an inhomogeneous electromagnetic field.

\subsection{Localized barriers I}

The Dirac Hamiltonian reads 
\begin{equation}\label{Rosen-Morse}
 H_M=i\sigma_1\partial_x+\left(A\tanh\,\alpha x+B\right)\sigma_2+M\sigma_3
\end{equation}
The operator with $M=0$ was discussed by Milpas in \cite{92} where the vector potential was considered as an exactly solvable approximation of the vector potential at the edge of the magnetized cobalt film \cite{492}. The Hamiltonian was utilized for the scattering analysis of the Dirac particles in graphene on the magnetic barrier. The exact solvability of $(H_M-E)\psi=0$ stems from the fact that $H_0$ corresponds to supercharge of the nonrelativistic, supersymmetric Rosen-Morse II Hamiltonian \cite{490}. The exact solvability of $(H_0^2-E)\psi=0$ is facilitated by the shape-invariance of the non-relativistic operator, see \cite{489}.

The potential terms in the transformed Hamiltonian $\tilde{H}=G^{\pm}H_{M}(G^{\pm})^{-1}=-i\sigma_1\partial_x+\tilde{V}_2^{\pm}\sigma_2+\tilde{m}\sigma_3+\tilde{V}_0^{\pm}$
acquire the following explicit form
\begin{eqnarray}\label{morse}
 \tilde{V}^{\pm}_0&=&\frac{A\, \alpha\, {\rm sech}^2\alpha x}{2\left(M^2+\left(B+A\tanh \alpha x\right)^2\right)}\left(M\pm\frac{\tilde{m} (B+A\tanh \alpha x)}{\sqrt{M^2-\tilde{m}^2+(B+A\tanh \alpha x)^2}}\right),\\
 \tilde{V}_2^{\pm}&=&\mp\sqrt{M^2-\tilde{m}^2+\left(B+A\tanh\alpha x\right)^2}.
\end{eqnarray}

\begin{figure}
\centering
\begin{tabular}{cc}
\includegraphics[scale=.6]{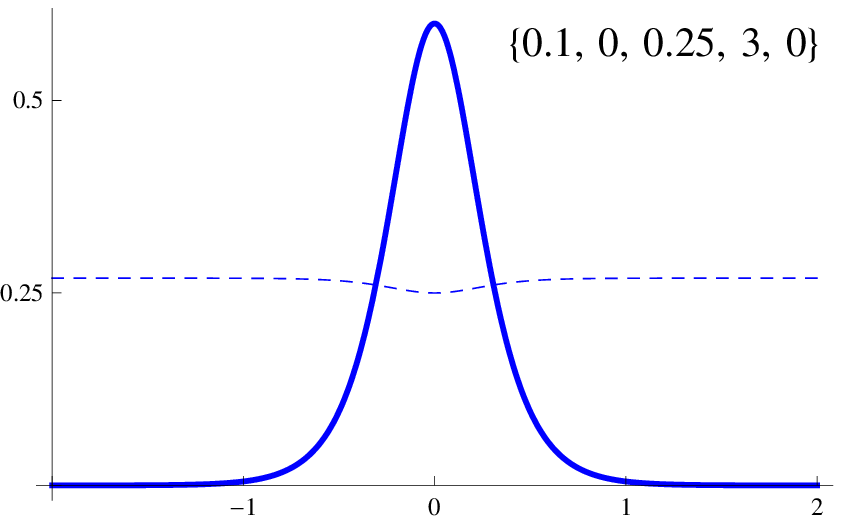}&\includegraphics[scale=.6]{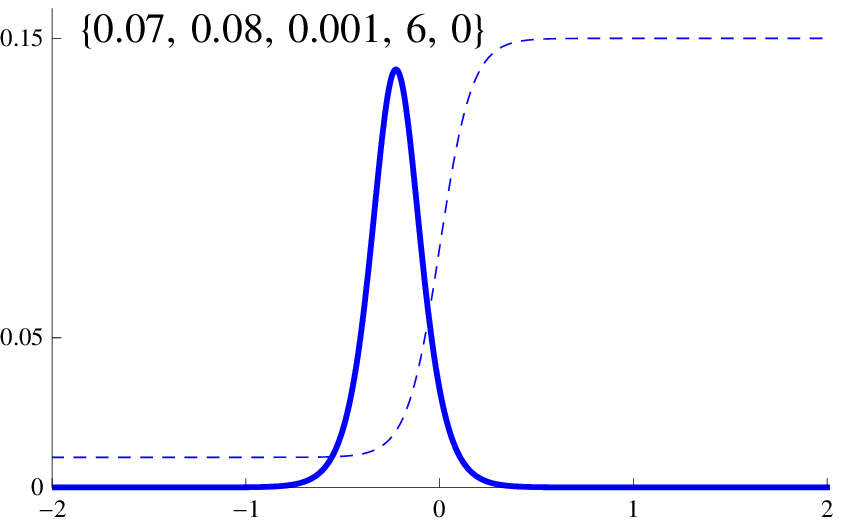}\\
a)&b)
\end{tabular}\caption{The potentials $\tilde{V}_0^{-}$ (thick line) and $\tilde{V}_2^-$ (dashed line). The values of parameters $\{A,B,M,\alpha,m\}$ are in the inset.}\label{RMII}
\end{figure}

The electrostatic potential $\tilde{V}_0^{-}$ decays exponentially, whereas $\lim_{x\rightarrow\pm \infty}\tilde{V_2}^-=\tilde{v}_2^{\pm}=\sqrt{M^2-\tilde{m}^2+\left(B\pm A\right)^2 }$. When we fix $B=0$, then $\tilde{v}_2^{+}=\tilde{v}_2^-$. We identify the non-zero asymptotic value of $V_2^-$ with $k_y$, whereas the asymptotically vanishing inhomogeneous component with the magnetic vector potential (see (\ref{tildeHH}) and related discussion). The Hamiltonian then describes Dirac fermion bouncing on the electrostatic (and a weak magnetic) barrier with the momentum $k_y=\sqrt{M^2-m^2+A^2 }$ parallel with the barrier, see Fig~\ref{RMII}a) for illustration. When $B>0$, $\tilde{V}_2^-$ represents a positive, smoothened step-like barrier, see Fig.~\ref{RMII}b). The Hamiltonian can be interpreted as describing Dirac fermion in radially twisted carbon nanotube where the asymptotically constant speed of twist gets changed by a local interaction with an anchor, represented by $\tilde{V}_0^-$, see \cite{85} and 
references therein. Qualitatively the same situation can be described by $\tilde{V}_0^+$ and $\tilde{V}_2^+$.

\subsection{Localized barriers II}
Let us fix the exactly solvable Hamiltonian in the following form
\begin{equation}
 H_M=i\sigma_1\partial_x+\left(A\tanh \alpha x+B\,\mbox{sech}\,\alpha x\right)\sigma_2+M\sigma_3,
\end{equation}
where the operator $H_0$ is known as the supercharge of the Scarf II supersymmetric Hamiltonian. The stationary equation $(H_0-E)\psi=0$ is exactly solvable due to the shape-invariance of $H_0^2$, see  \cite{489}. Hence, the equation $(H_M-E)\psi=0$ is exactly solvable as well. The gauge transformed Hamiltonian  $\tilde{H}=G^{\pm}H_M(G^{\pm})^{-1}$ has the following potential terms
\begin{eqnarray}
 \tilde{V}_0^{\pm}&=&\frac{\alpha\,{\rm sech^2}\alpha\, x(A-B{\rm sinh} \alpha \,x)\left(M\pm\frac{\tilde{m}\,{\rm sech}\alpha \,x(B+A\sinh \alpha \,x)}{\sqrt{M^2-\tilde{m}^2+(B{\rm sech}\, \alpha \,x+A\tanh \alpha \,x)^2}}\right)}{2(M^2+(B{\rm sech}\,\alpha x+A\tanh\, \alpha \,x)^2)}\\
 \tilde{V}_2^{\pm}&=&\mp\sqrt{M^2-\tilde{m}^2+(B{\rm sech}\,\alpha \,x+A\tanh \alpha \,x)^2}.
%\\ \tilde{V_3}&=&m.
\end{eqnarray}

\begin{figure}
\centering
\begin{tabular}{cc}
\includegraphics[scale=.6]{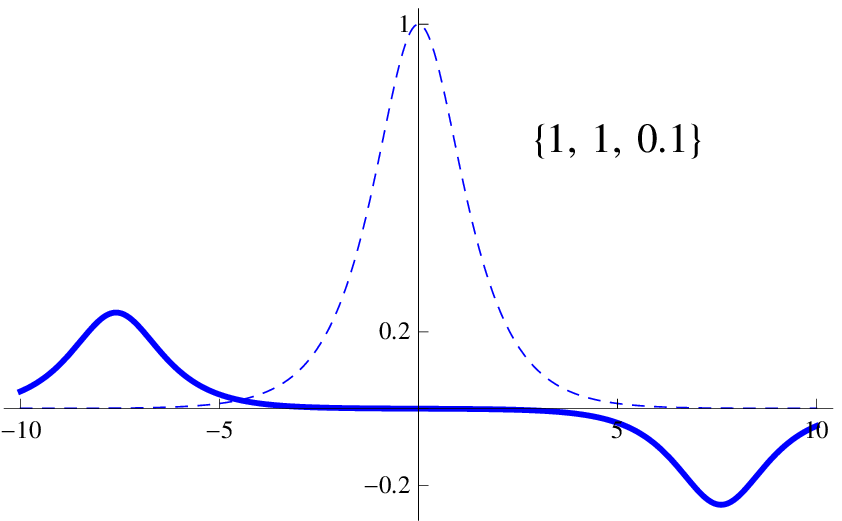}&
\includegraphics[scale=.6]{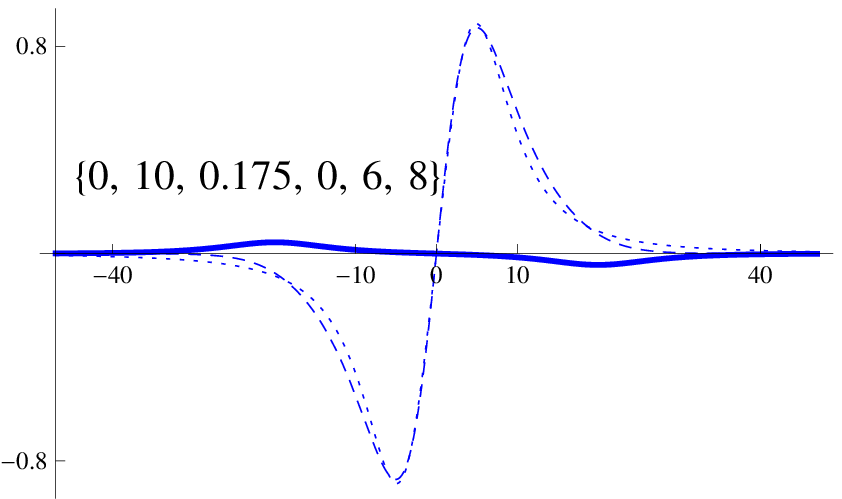}\\
a)&b)
\end{tabular}\caption{a) $\tilde{V}_2^--M$ (dashed line), $\tilde{V}_0^-$ (solid line) from (\ref{scarfspec}) for $\{B,\alpha,m\}$ specified in the inset. b) $\tilde{V}_0^+$ (thick line), $(\tilde{V}_2^+)'$ (dashed line), $B_z$ from (\ref{Bz}) (dotted line) for $B_0=1$ and $\{A,B,\alpha,m,z_0,d\}$, see the inset.}\label{figscarf}
\end{figure}

When $A=0$ and $\tilde{m}=0$, the electrostatic potential acquires a simplified form
\begin{equation}\label{scarfspec}
 \tilde{V}_0=-\frac{B M\alpha\,{\rm sech}\, \alpha x\, {\rm tanh}\, \alpha x}{2(M^2+B^2{\rm sech}^2\, \alpha x)},\quad \tilde{V}^{\pm}_2=\mp \sqrt{M^2+B^2{\rm sech^2}\,\alpha x}.
\end{equation}
The vector potential $\tilde{V}_2$ is asymptotically constant. We assign its  asymptotic value to the nonvanishing momentum $k_y=M$ and its inhomogeneous part to the external magnetic vector potential.
The scalar potential $\tilde{V}_0$ represents two bumps situated at $\alpha x=\pm{\rm arccosh}\sqrt{\frac{B^2+2M^2}{M^2}}$, see Fig.\ref{figscarf}~a). When $M\rightarrow 0$, the distance $2\alpha^{-1}{\rm arccosh}\sqrt{\frac{B^2+2M^2}{M^2}}$ between the bumps goes to infinity while $k_y$ in $\tilde{V}_2$ tends to zero. 

The model can serve as a good approximation of the field created by a a thin, infinitely long strip  of ferromagnetic material situated along $y$-axis, of the width $d$, and in the distance $z_0$ from the graphene sheet. The vector of magnetization is parallel to $x$-axis. Magnetic field induced by such strip was calculated in \cite{491} and reads
\begin{equation}\label{Bz}
 B_z=B_0\left(\frac{z_0 d}{\left(x-\frac{d}{2}\right)^2+z_0^2}-\frac{z_0 d}{\left(x+\frac{d}{2}\right)^2+z_0^2}\right).
\end{equation}
This profile can be matched quite well by the vector potential $\tilde{V}_2-M$, see Fig.\,\ref{figscarf}~b).

\subsection{Spectral isomorphism between two- and three-gap periodic systems}

Analysis of Dirac particles in graphene in periodic fields is quite topical. Graphene superlattices can be prepared experimentally 
by putting the graphene sheet onto a metallic substrate 
\cite{515}, by thin films of ferromagnetic material 
\cite{507} or by a periodically structured substrate that induces strains in the graphene sheet \cite{511}. 
Solvable models describing superlattices were considered in the literature, e.g. the barriers of Kronig-Penny type \cite{91}, periodic step-like potentials  \cite{12},  or sinusoidal barriers \cite{198}.

Let us take the following Hamiltonian
\begin{equation}
 H_M=-i\sigma_1\partial_x- k^2\frac{sn( x,k)cn( x,k)}{dn( x,k)}\sigma_2+M\sigma_3
\end{equation}
as the initial operator. 
Here, $sn(x,k)$, $cn(x,k)$ and $dn(x,k)$ are Jacobi elliptic functions. It is worth mentioning that the vector potential of $H_0$ was considered as a self-consistent crystalline condensate in \cite{156}. The Hamiltonian $H_M$ is $2K(k)$ periodic ($K(k)$ is the complete elliptic integral of the first kind and $k\in\langle 0,1\rangle$ is the modular parameter), so that its spectrum has band structure. The peculiar properties of $H_M$ originate from the fact that $H_0$ belongs to the family of finite-gap operators; $H_0$ has just two gaps in its spectrum. 

The solution of the stationary equation for $H_0$ can be deduced easily from the solutions of the one-gap Lam\'e equation ($(H_0^2-E)\psi=0$) that are well known \cite{Whittaker}. Let us define the spin-up spinors
\begin{equation}
 \Xi_{\pm}=\left(\frac{\Theta_1\left(\frac{\pi(x\pm\alpha)}{2K(k)}\right)}{\Theta_4\left(\frac{\pi x}{2K(k)}\right)},0\right)^Te^{\mp \xi(\alpha,k)}.
\end{equation}
Here $\Theta_1$, $\Theta_4$ and $\zeta$ are Jacobi theta functions and Zeta function, respectively, see \cite{Whittaker}. The parameter $\alpha$ can acquire any complex value in principle. However, $\Xi_{\pm}$ are quasi-periodic Bloch states if and only if \footnote{Modulo periodicity of $dn(\alpha,k)$.} $\alpha=i\delta$ or $\alpha=K(k)+i\delta$, where $\delta\in \langle0,2K(k')\rangle$ and $k'^2=1-k^2$. In correspondence with these allowed values of $\alpha$, the operator $H_0$ has three bands of energies, $E\in(-\infty,-1\rangle\cup\langle -k',k'\rangle\cup\langle 1,\infty)$. 
The eigenfunction of $H_0$ for $E\geq0$ (they correspond to $\alpha\in  i\langle0,K(k')\rangle$ or $\alpha\in K(k)+i\langle0,K(k')\rangle$) can be constructed as
\begin{equation}
 \Psi_{\pm}^0=\Xi_{\pm}+\frac{1}{E}H_0\Xi_{\pm},\quad H_0\Psi^0_{\pm}=E\Psi^0_{\pm},\quad E=dn(\alpha,k),
\end{equation}
whereas the eigenstates for $E<0$ can be obtained as $\sigma_3\Psi_{\pm}^0$.

In accordance with (\ref{psiM}), we can write the wave functions of $H_M$ corresponding to the positive energy $E>M$ in the following form
\begin{equation}
\Psi^{\pm}_{|\epsilon|}=\Xi_{\pm}+\frac{1}{|\epsilon|+M}H_0\Xi_{\pm},\quad H_M\Psi^{\pm}_{|\epsilon|}=|\epsilon|\Psi^{\pm}_{\epsilon},
\end{equation}
whereas the wave functions for $E<-M$ read 
\begin{equation}
\Psi^{\pm}_{-|\epsilon|}=\Xi_{\pm}+\frac{1}{|\epsilon|-M}\sigma_3H_0\Xi_{\pm},\quad H_M\Psi^{\pm}_{-|\epsilon|}=-|\epsilon|\Psi^{\pm}_{-|\epsilon|}.
\end{equation}
The spectrum of $H_M$ consists of the following four intervals 
$ \sigma(H_M)=\left(-\infty,-\sqrt{1+M^2}\right\rangle\cup\left\langle-\sqrt{k'^2+M^2},-M\right\rangle\cup\left\langle M,\sqrt{k'^2+M^2}\right\rangle\cup\left\langle\sqrt{1+M^2},\infty\right).$
There are six band-edge states. They are
\begin{equation}
 \Psi_M=(dn x,0)^T,\quad \Psi_{-M}=\left(0,dn(x+K)\right)^T,
\end{equation}
\begin{equation}
\Psi_{\pm\sqrt{k'^2+M^2}}=\left(\begin{array}{r}cn(x)\\\mp i\frac{\sqrt{k'^2}}{\sqrt{k'^2+M^2}\pm M}cn(x+K)\end{array}\right),\quad 
\Psi_{\pm\sqrt{1+M^2}}=\left(\begin{array}{r}sn(x)\\\mp i\frac{1}{\sqrt{1+M^2}\pm M}sn(x+K)\end{array}\right).
\end{equation}
\begin{figure}
\centering
\begin{tabular}{cc}
\includegraphics[scale=.6]{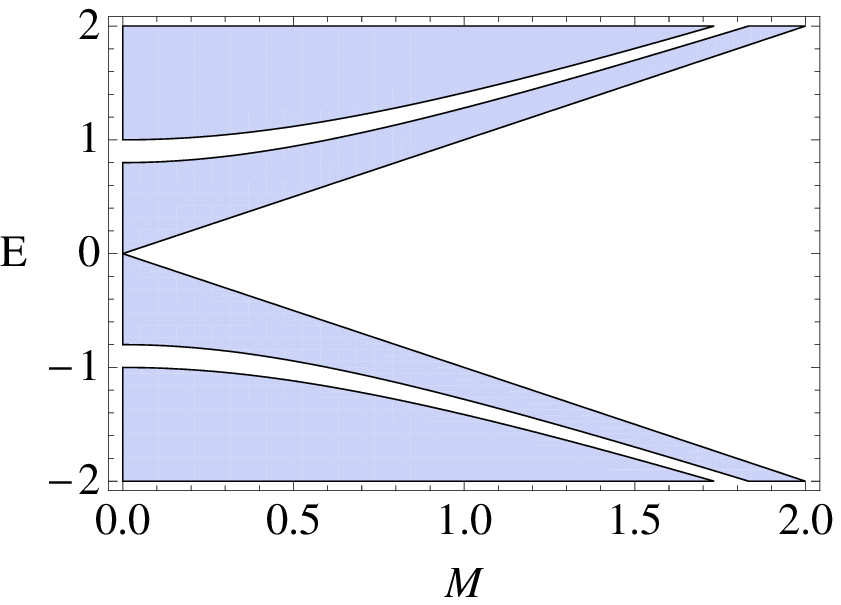}&\includegraphics[scale=.6]{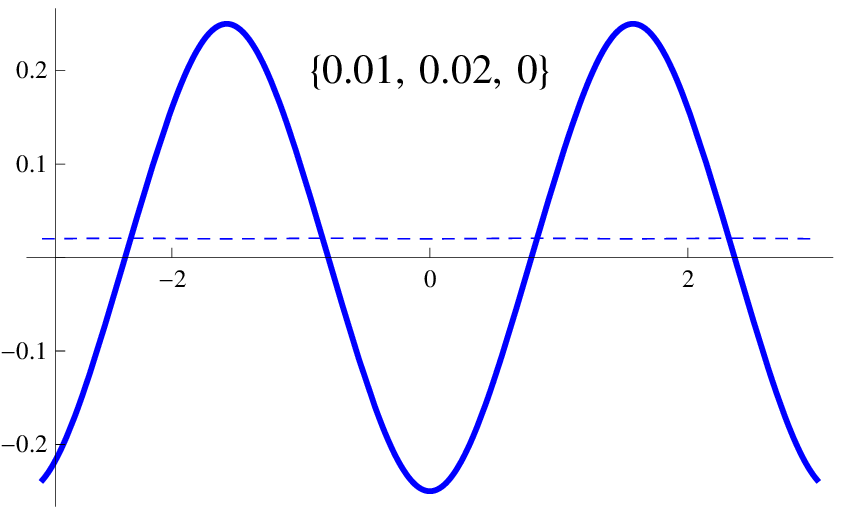}\\
a)&b)\\
\includegraphics[scale=.7]{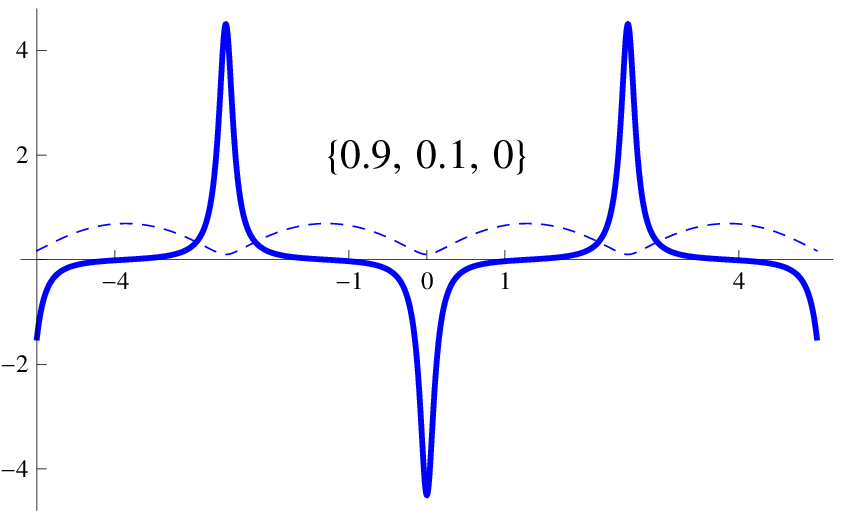}&\includegraphics[scale=.6]{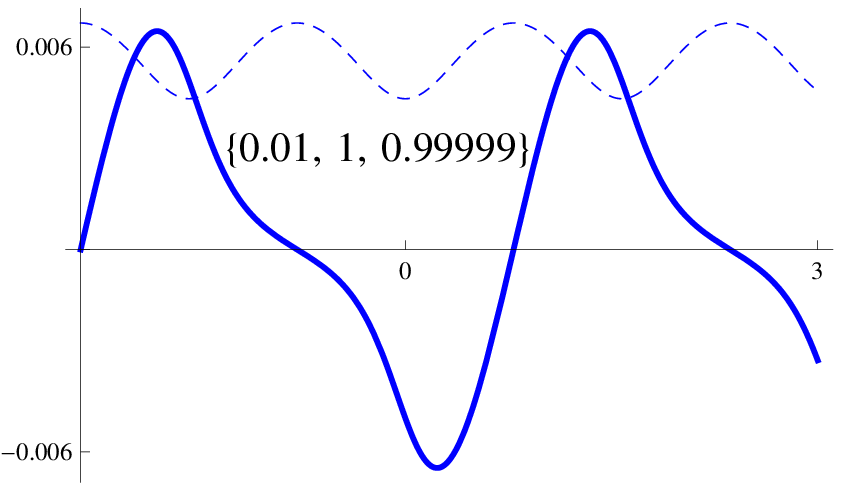}\\
c)& d)
\end{tabular}\caption{a) the vertical axis corresponds to $\sigma(H_M)$, the horizontal axis to $M$, $k=0.6$. Shadowed regions represent allowed energies. b)-d) $\tilde{V}_2^-$ dashed line, $\tilde{V}_0^-$ solid line. The values of parameters $\{k^2,M,\tilde{m}\}$ are in the inset of each plot. }\label{lame1}
\end{figure}

Addition of the mass term into the two gap Hamiltonian $H_0$ opens an additional central gap in the spectrum of $H_M$ that corresponds to the three-gap system, see Fig. \ref{lame1}a). This is underlined by the fact that its potential term solves the corresponding nonlinear equation of the Ablowitz-Kaup-Newell-Segur (AKNS) hierarchy. Defining $\Delta_M=ik^2\frac{sn x cn x}{dn x}-M$, it solves 
\begin{equation}\label{AKNS}
 AKNS_3=\frac{1}{8}\left(-\Delta_M^{(3)}+6 |\Delta_M|^2\Delta_M'\right)+\frac{1}{4}\left(-2+k^2-3M^2\right)\Delta_M'=0,
\end{equation}
\begin{equation}\label{AKNS0}
 AKNS_2=-\frac{\Delta_0''}{4}+\frac{1}{2}|\Delta_0|^2\Delta_0+\frac{1}{2}(k^2-2)\Delta_0=0,
\end{equation}
where $(AKNS_2)'\sim AKNS_3$ for $M=0$. Let us notice that both the two-gap Hamiltonian $H_0$ and the three-gap operator $H_M$ have a nontrivial integrals of motion represented by a higher-order differential operators with matrix coefficients. Their existence emerges as a consequence of (\ref{AKNS}), see e.g. \cite{517} for details on their generic construction.
 
The unitary transformation allows to construct Dirac Hamiltonian with the oscillating electromagnetic field. The transformation (\ref{gauge}) is well defined (the functions $g_0^{\pm}$ and $g_1^{\pm}$ are continuous and real) since we take $\tilde{m}<V_3=M$ and $V_2$ is bounded. The Hamiltonian $\tilde{H}$ has the following potential terms
\begin{eqnarray}\label{Glame}
 \tilde{V}_0^{\pm}&=&\frac{1-k^2-dn^4 x}{2dn^2x\left(M^2+k^4\frac{cn^2xsn^2x}{dn^2x}\right)}\left(M\mp \frac{k^2 \tilde{m} \frac{cn x sn x}{dn x}}{ \sqrt{M^2-\tilde{m}^2+k^4\frac{cn^2xsn^2x}{dn^2x}}}\right),\\
 \tilde{V}_2^{\pm}&=&\mp\sqrt{M^2-\tilde{m}^2+k^4\frac{cn^2xsn^2x}{dn^2x}}.
\end{eqnarray}
In dependence on the parameters $M$, $k$, and $\tilde{m}$, the two potentials can acquire remarkable variety of profiles, see Fig.~\ref{lame1}b)-d). For instance, the vector potential can exhibit only mild oscillations around a non-zero constant that are negligible in comparison to the amplitude of the electrostatic potential, see Fig.~\ref{lame1}b). This system can be seen as an approximation of the setting where Dirac particle travels in the $2D$ periodic array of electrostatic barriers with the momentum $k_y\equiv \tilde{V}_2$ parallel to the barriers. The small, periodic fluctuations of $\tilde{V}_2$ in our model facilitate solution of the stationary equation that wouldn't be analytically solvable for constant $\tilde{V}_2=k_y$. When $\tilde{m}\neq 0$, the reflection symmetry of the scalar potential $V^{\pm}_0$ is spoiled, see Fig.~\ref{lame1}d). Instead, the potential seems to be odd with respect to the  reflection within the finite intervals determined by two consecutive extrema of the potential.

\section{Discussion}
In the article, we focused on the construction of solvable Dirac operators with inhomogeneous vector and scalar potential, and constant (possibly vanishing) mass. The core of our results represents the unitary transformation (\ref{G}), (\ref{gauge}) that brings a generic one-dimensional Dirac operator into the required form. We illustrated its application on the special class of solvable models that are based on the mass-less, solvable Dirac operator with pure vector potential. We showed that solvability of these systems is not compromised by addition of the mass term, yet the spectrum changes globally and gives rise to the spectral isomorphism (\ref{Ee}) between the systems. 
The constructed models were used in description of Dirac fermions in graphene in presence of an inhomogeneous electromagnetic field. 

We presented solvable models where the particle experiences localized as well as periodic electromagnetic barrier. The two settings with localized barrier proved to be good approximation for magnetic field induced either on the edge of the magnetized film (\ref{morse}) or by a magnetized strip of finite width (\ref{scarfspec}). In case of the periodic array of barriers, the system with two gaps in the spectrum served as the initial ``seed'' system. We constructed a spectrally isomorphic Hamiltonian with three spectral gaps and discussed some its peculiar properties, e.g. its relation to the AKNS hierarchy of integrable systems (\ref{AKNS}).   

The presented procedure is complementary to the existing methods, namely to the Darboux transformation \cite{474}. The latter one can transform an initial Dirac operator into an (almost) isospectral Hamiltonian with altered vector potential and mass term. The presented unitary transformation can set the mass to a constant and induce an inhomogeneous scalar potential. It can provide the new isospectral partners of the recently discussed Dirac systems with reflection-less potential \cite{538}, \cite{reflectionDirac} or inhomogeneous magnetic fields \cite{429}, \cite{alonso}. 
It is worth noticing that a different approach to construction of solvable systems described by Dirac equation was developed in \cite{419}. There, a transformation involving analytic continuation of parameters was discussed that allowed to get new solvable Dirac operators from the known ones with, in  specific manner interchanged, vector and scalar potentials.

It would be interesting to consider spectral isomorphism for generic finite-gap systems. In particular, it should be understood how the procedure, which maps an $n-$gap Hamiltonian to an $(n+1)$-gap one, is reflected  by the associated equations of the AKNS hierarchy, or manifested in the structure of  (Lax) integrals of motion associated with the finite-gap Hamiltonians. In this context, the results on of the hidden nonlinear supersymmetry of the finite-gap systems could be particularly useful \cite{517}.

As the operator $\tilde{H}^2$ represents a Schr\"odinger operator with matrix potential (the antidiagonal terms with first derivative can be transformed out be an additional unitary transformation), the presented procedure can be beneficial for analysis of these nonrelativistic systems that are of growing interest recently \cite{397}, \cite{sokolov1}, \cite{reflectionDirac}.

The presented construction was primarily aimed at derivation of explicit exactly solvable models. However, it can be utilized for qualitative analysis as well. The Hamiltonians $H_M$ were discussed qualitatively in \cite{502}, where a set of sufficient conditions for existence of bound states was derived. With the use of the unitary transformation (\ref{gauge}), those qualitative results are directly applicable on a class of operators $\tilde{H}$ with inhomogeneous vector and scalar potential.


\begin{thebibliography}{10}
\bibitem{NJL} 
  Y.~Nambu and G.~Jona-Lasinio,
 % ``Dynamical Model of Elementary Particles Based on an Analogy with Superconductivity. 1.,''
  Phys.\ Rev.\  {\bf 122}, 345 (1961).
 %%CITATION = PHRVA,122,345;%%
  %3732 citations counted in INSPIRE as of 30 Apr 2013
  
\bibitem{152} 
D. J. Gross, A. Neveu, %`` Dynamical symmetry breaking in asymptotically free field theories,''
Phys.\ Rev.\ D {\bf 10}, 3235 (1974).

\bibitem{519}
M. Buballa, S. Carignano, `` Inhomogeneous chiral condensates,'' arxiv:1406.1367.

\bibitem{520}  A. F. Andreev, %``The Thermal Conductivity of the Intermediate State in Superconductors,'' 
Soviet Physics JETP {\bf 19}, 1228 (1964).

\bibitem{345}G.\ W.\ Semenoff, %``Condensed-Matter Simulation of a Three-Dimensional Anomaly,''
Phys.\ Rev.\ Lett.\ {\bf 53}, 2449 (1984).

\bibitem{84}V. Jakubsk\'y, L. M. Nieto, M. S. Plyushchay, %``Klein tunneling in carbon nanostructures: A free-particle dynamics in disguise,''
Phys.\ Rev.\ D {\bf 83}, 047702 (2011).

\bibitem{529}A. A. Andrianov, F. Cannata, %`` Nonlinear supersymmetry for spectral design in quantum mechanics,''
J.\ Phys.\ A {\bf 37},  10297 (2004).

%\bibitem{0-ordersusy}Due to its formal similarity with the (non-linear) supersymmetric quantum mechanics, the unitary transformation that intertwines two Hamiltonians that correspond to distinct physical situations ($UH_0=\tilde{H}U$) was coined as a supercharge of the $0$-order supersymmetry in \cite{84} 

\bibitem{474}L. M. Nieto, A. A. Pecheritsin, B. F. Samsonov, %``Intertwining technique for the one-dimensional stationary Dirac equation,''
Annals of Physics {\bf 305}, 151 (2003).

\bibitem{85}V. Jakubsk\'y, M.\ S.\ Plyushchay, %``Supersymmetric twisting of carbon nanotubes,''
Phys.\ Rev. D {\bf 85}, 045035 (2012).

\bibitem{97}M. A. H. Vozmediano, M. I. Katsnelson, F. Guinea,  %``Gauge fields in graphene,''
Phys.\ Rep.\ {\bf 496}, 109 (2010).

\bibitem{10}B. Hunt et al, %`` Massive Dirac fermions and Hofstadter butterfly in a van der Waals heterostructure,''
Science {\bf 340}, 1427-30 (2013).

\bibitem{250}P. M. Krstajic, P. Vasilopoulos, J.\ Phys.\ Cond.\ Matt.\ {\bf 23}, 135302 (2011).

\bibitem{92} E.\ Milpas, M.\ Torres and G.\ Murgu\'i{}a, J.\ Phys.\ Condens.\ Matter\ {\bf23}, 245304 (2011)

\bibitem{492}T. Van\v cura et al %``Electron transport in a two-dimensional electron gas with magnetic barriers,''
Phys.\ Rev.\ B {\bf 62},  5074 (2000).

\bibitem{490}N.\ Rosen, P.\ M.\ Morse, Phys.\ Rev.\ {\bf 42}, 210 (1932).

\bibitem{489}R. Dutt, A. Khare, U. P. Sukhatme, Am.\ J.\ Phys.\ {\bf 56}, 163 (1988).

\bibitem{491}A.\ Matulis, F.\ M.\ Peeters, P.\ Vasilopoulos,
%``Wave-vector-dependent tunneling through magnetic barriers'' 
Phys.\ Rev.\ Lett.\ {\bf 72}, 1518  (1994).

%\bibitem{514}J. C. Meyer, C. O. Girit, M. F. Crommie, A. Zettl, % ``Hydrocarbon lithography on graphene membranes''
%Applied Physics Letters {\bf 92} 123110 (2008).   

\bibitem{515} D. Martoccia et al, % ``Graphene on Ru(0001): A 25×25 Supercell,''
Phys.\ Rev.\ Lett.\ {\bf 101}, 126102 (2008).

%\bibitem{2}W. van Roy, %``Study of the demagnetization and optimization of the magnetic field of perpendicular ferromagnetic thin films usingesub-μ m lithography,''
%Journal of Magnetism and Magnetic Materials {\bf 121}, 197-200 (1993).

\bibitem{507}P. Ye, D. Weiss, R. Gerhardts, M. Seeger, K. von Klitzing, K., K. Eberl, H. Nickel, %``Electrons in a Periodic Magnetic Field Induced by a Regular Array of Micromagnets,'' 
Phys.\ Rev.\ Lett.\ {\bf 74}, 3013 (1995). 

\bibitem{511}A. V\'azquez de Parga et al, %``Periodically Rippled Graphene: Growth and Spatially Resolved Electronic Structure,''
Phys.\ Rev.\ Lett.\ {\bf 100}, 056807 (2008).

%\bibitem{95}M. Ramezani, P. Vasilopoulos, F. M. Peeters, %``Magnetic Kronig–Penney model for Dirac electrons in single-layer graphene,''
%New Journal of Physics, {\bf 11}, 095009 (2009).

%\bibitem{505}V. Qui Le, C. Huy Pham, V. Lien Nguyen, %``Magnetic Kronig–Penney-type graphene superlattices: finite energy Dirac points with anisotropic velocity renormalization,''
%J. Phys. Condens Matter {\bf 24}, 345502 (2012).

\bibitem{91}M. Ramezani, P. Vasilopoulos, F. M. Peeters, %``Kronig-Penney model of scalar and vector potentials in graphene,''
J.\ Phys.\ Condens.\ Matter {\bf 22}, 465302 (2010).

\bibitem{12}L. Dell'Anna, A. de Martino, %``Multiple magnetic barriers in graphene,''
Phys.\ Rev.\ B {\bf 79}, 45420 (2009).

%\bibitem{512}M. Barbier, P. Vasilopoulos, F. M. Peeters, %``Extra Dirac points in the energy spectrum for superlattices on single-layer graphene,''
%Physical Review B {\bf 81}, 075438 (2010).

\bibitem{198}L. Brey, H. A. Fertig, %``Emerging Zero Modes for Graphene in a Periodic Potential'', 
Phys.\ Rev.\ Lett.\ {\bf 103}, 046809 (2009).

%\bibitem{510}Park, Cheol-Hwan; Zheng Tan, Liang; Louie, Steven G., %``Theory of the electronic and transport properties of graphene under a periodic electric or magnetic field,''
%Physica E {\bf43}, 651-656 (2011).

%\bibitem{506} Cheol-Hwan Park, L. Yang, Y.-W. Son, M. L. Cohen, S. G. Louie, %`` Anisotropic behaviours of massless Dirac fermions in graphene under periodic potentials,''
%Nature Physics {\bf 4}, 213 (2008).

%\bibitem{206}I. Snyman, %``Gapped state of a carbon monolayer in periodic magnetic and electric fields,''
%Physical Review B {\bf 80}, 054303 (2009).

\bibitem{156}
G.\ Basar, G. V. Dunne, %``Twisted kink crystal in the chiral Gross-Neveu model,''
Phys.\ Rev.\ D {\bf 78}, 065022 (2008).

\bibitem{Whittaker}E.\ T.\ Whittaker, G.\ N.\ Watson, A Course of Modern Analysis, Cambridge University Press, p.573

\bibitem{517}F. Correa, G. V. Dunne, M. S. Plyushchay, %``The Bogoliubov–de Gennes system, the AKNS hierarchy, and nonlinear quantum mechanical supersymmetry,''
Annals of Physics {\bf 324}, 2522 (2009).

\bibitem{538}A.~Arancibia, M.~S.~Plyushchay, %``Transmutations of supersymmetry through soliton scattering and self-consistent condensates,''
Phys.\ Rev.\ D {\bf 90}, 025008 (2014).

\bibitem{reflectionDirac} 
  F.~Correa and V.~Jakubsk\'y,
  ``Twisted kinks, Dirac transparent systems and Darboux transformations,''
  arXiv:1406.2997 [hep-th].
  %%CITATION = ARXIV:1406.2997;%%
  %2 citations counted in INSPIRE as of 04 Nov 2014
  
\bibitem{429}B.~Midya, D. J. Fern\'andez, J. Phys. A {\bf 47}, 285302 (2014).

\bibitem{alonso}A. Contreras-Astorga, A. Schulze-Halberg, J.\ Math.\ Phys.\ {\bf 55}, 103506 (2014).

\bibitem{419}L. Z. Tan, Cheol-Hwan Park, S. G. Louie, %``Graphene Dirac fermions in one-dimensional inhomogeneous field profiles: Transforming magnetic to electric field,''
Phys.\ Rev.\ B {\bf 81}, 195426 (2010).

\bibitem{397}A.~G.~Nikitin, Y.~Karadzhov, %``Matrix superpotentials,''
J.\ Phys.\ A {\bf 44},  5204 (2011).

\bibitem{sokolov1}  A.~V.~Sokolov, Phys.~Lett.~A {\bf377}, 655 (2013);\\ A.~A.~Andrianov, A.~V.~Sokolov, ``Minimal Realizations of Supersymmetry for Matrix Hamiltonians,'' arXiv:1409.6695 %A.~V.~Sokolov,  [arXiv:1406.0191].

%\bibitem{sokolov2} 

\bibitem{502}V.\ Jakubsk\'y, D.\ Krej\v ci\v r\'i{}k, %``Qualitative analysis of trapped Dirac fermions in graphene,''
Annals of Physics {\bf 349}, 268 (2014).






















  


%\cite{Nambu:1961tp}









\end{thebibliography}
\end{document}